

\vglue 2.5 cm
\centerline {\bf Optimal photon localization}
\vskip 1 cm

\centerline {J. Mourad }
\centerline {Laboratoire de Physique Th\'eorique et Hautes Energies\footnote*
{Laboratoire associ\'e au CNRS}}
\centerline {Bat. 211, Universit\'e Paris-Sud, 91405 Orsay, France.}
\vskip 1 cm
\noindent
{\bf Abstract :} We show that there is only one operator
having some minimal properties
enabling
it to be a one photon position operator. These properties are stated, and
the solution is shown to be the photon position operator proposed by Pryce.
This operator has non-commuting components. Nevertheless, it is shown
that one can find states localized with an arbitrary precision.

\vfill
\eject
\beginsection {1.Introduction}

The experimental possibility of preparing one photon states greatly motivates
the question of the photon localizability [1]. The question we wish to answer
is the
following: suppose we know the wave function of the photon, can we calculate
the probability of finding it somewhere in space at a particular time ?

Different answers have been given to this question. Newton and Wigner [2] and
later on Wightman [3] argued that the answer is negative, while Jauch, Piron
[4]
 and
Amrein [5] gave a positive answer at the expense of changing the axioms of
localizability.( They essentially dropped the requirement that the probability
of
finding a particle in a space region $R$ is the sum of the probabilities of
finding it in the disjoint sub-regions $R_i$, where $\bigcup_{i} R_i=R$ ).
The axioms of Newton and Wigner were for exactly localized sates, they
therfore assumed that if no such states exist the particle is not localizable.
In other words, they proved that no position operator with commuting components
exists for the photon.

In this letter we  drop the assumption of commuting
components and retain the other Newton-Wigner axioms. We  show, in section 2,
that there exists an operator obeying these axioms and that this operator
is unique. This operator was found for the first time by Pryce while
looking for the analogue of the centre of mass for relativistic wave equations
[6,7].
We will call it the Pryce operator.

Next we show, in section 3, that although exactly localized states do not
exist,
one can find
states that are localized within an arbitrary small region of
non vanishing volume.
We argue that the Pryce
operator can describe the position of the photon and gives means to answer the
question we asked at the beginning.

 \beginsection {2.The position operator}

 The photon is described by two irreducible representations
 of the Poincar\'e group [8], one for each helicity state. The minimal
 description of these representations is by means of
 a two-component wave function. However, this description is singular, the
 generators of the Poincar\'e group having Dirac-like monopole
 singularities, and the wave functions being elements of an associated Hilbert
 space over a U(1) -fiber bundle [9]. Another description,
 equivalent to the previous one
 is obtained by taking the zero mass limit of the massive representation
 of a spin one particle and constraining it to have no zero helicity
 components. This
 description is free of singularities and we will use it in the following.
 We choose the scalar product:
 $$ (\phi,\psi)= \sum_{i=1}^3 \int d{\bf p} \phi^*_i({\bf p}) \psi_i({\bf p})
 .\eqno (2.1) $$
 The generators of the Poincar\'e group are unitarily equivalent to
 $${\bf P}={\bf p} \ ,\  \ H=\sqrt {\left({ {\bf p}^2 }\right)},\eqno (2.2a)$$
 $$ {\bf J}=i{\partial \over { \partial {\bf p}}} \wedge { \bf p } + {\bf s}
 ,\ \  {\bf K}={i \over 2} \{ {\partial \over {\partial {\bf p}}},H \}
 -{{\bf s}\wedge {\bf p} \over H}. \eqno (2.2b)$$
 Here $ \bf P$,$H$,$\bf J$ and $\bf K$ are the generators of space
translations,
 time evolution and boosts, $\bf s$ are the three dimensional generators
 of the $SO(3)$ group. The parity and time inversion operators, $\Pi$ and
 $\Theta$
 act on the wave function as follows
 $$ \Pi \psi_i ({\bf p})= \psi_i (-{\bf p}), \ \ \Theta \psi_i({\bf p})
 =\tau _{ij}\psi^*_j(-{\bf p}).\eqno (2.3)$$
 The matrix $\tau$ is defined by the equation
 $$ \tau ^{ \dag} s  \tau=-s ^{*}.\eqno (2.4)$$

 The wave function, in addition, is constrained to have a vanishing
 zero-helicity component
 $$   { ({\bf p.s})^2 \over {{\bf p}^2}}\psi=\psi.\eqno (2.5)$$
 The operator $({\bf p.s})^2 \over {{\bf p}^2}$ is the projector
 onto non vanishing helicity states. note that it commutes
 with the generators of the Poincar\'e group,as it should.
 Now that the photon description is complete we have to define
 the position operator. We will do this by requiring that it
  satisfies certain properties and then look for operators that do obey
 these requirements.

 The position operator eigenvalues must be real so the operator
 must be self-adjoint
 $$ {\bf x}= {\bf x}^{\dag}.\eqno (2.6)$$
 It must transform under  space translations and rotations as
  a vector
 $$ e^{i {\bf P.a}} {\bf x} e^{-i {\bf P.a}}=
 {\bf x} + {\bf a}, \  \ e^{i{\bf J.n}} {\bf x}_i e^{-i{\bf J.n}}=R({\bf
n})_i^j
 {\bf x}_j .\eqno (2.7)$$
 Here R(n) is the three dimensional rotation matrix.
 Taking infinitesimal transformations in equations $(2.7)$, we find the
commutation
 relations

 $$[x_i,p_j]=i\delta_{ij},\ \ [x_i,J_j]=i\epsilon _{ijk}x_k.\eqno (2.8a)$$
  Lastly we require
 the position operator to transform as the classical position
 under parity and time inversion transformations
 $$ \Pi {\bf x} \Pi =-{\bf x},\ \ \Theta {\bf x} \Theta ^{-1}={\bf x}.
 \eqno (2.8b)$$
 In addition the position operator must act on physical
 states, those satisfying the constraint $(2.5)$, and the result must
 also be a physical state, so we have to impose the constraint
 $$ [{\bf x},{({\bf p.s})^{2} \over {{\bf p}^2}}]=0.\eqno (2.8c)$$
 Equations $(2.8)$ and $(2.6)$ are thus the properties we demand from a
position
 operator. We will show that there exists only one operator
 satisfying these equations.
 Equations $(2.8a)$ are solved as follows
 $$ {\bf x} =i{{{\partial} \over {\partial {\bf p}}}} +{\bf f},\eqno (2.9)$$
 where ${\bf f}$ is a vector constructed out of the momentum and spin
 operators. There are three basic vectors, ${\bf p},{\bf s}$ and
 ${\bf p \wedge s}$. The operator ${\bf f}$ is a linear combination of these
 vectors with scalar coefficients. The scalars may be constructed
 out of functions of ${\bf p}^2$ and of $\bf p.s$. The latter is actually
 a pseudoscalar, so it can only be the coefficient of ${\bf s}$; this
combination
 does not have the correct behavior under time inversion, so it must be
 dropped. Note that, due to the constraint (2.5), higher powers of ${\bf p.s}$
 can be dropped too.
  The same line of argument can be used to eliminate the term
 proportional to ${\bf p}$ so that we are left with the solution to equations
 $(2.6),(2.8a)$ and $(2.8b)$
 $$ {\bf f}=g({\bf p}^2)\ {\bf p \wedge s},\eqno (2.10)$$
 where g is an arbitrary function of ${\bf p}^2$. In order to have the correct
 dimension it must be proportionnal to $1\over {{\bf p}^2}$, the remainig
 coefficient being determined by the constraint $(2.8c)$ and equal to  1.
 Finally we get the unique solution
 $$ {\bf x}=i{\partial \over {\partial {\bf p}}} +{{\bf p \wedge s}
   \over {\bf p}^2 }.\eqno (2.11)$$
   This operator has been proposed for the first time by Pryce [6] when looking
   for the analogue of the centre of mass for the Maxwell equation.
   It may be written in a representation independent way
   $${\bf x}={1 \over 2} \{ {1 \over H},{\bf K} \}.\eqno (2.12)$$

The mean value of the Pryce position operator
in a state described by the wavefunction $\psi$ can be obtained by
the formula
$$ <{\bf  x}>= \int d{\bf r}\hat \psi^{\dag}({\bf r}){\bf r}\hat \psi({\bf r})
+ {i \over {4\pi}} \int d{\bf R}d{\bf u} \hat
 \psi^{\dag}({\bf R}+{{\bf u}\over {2}}
){{\bf u \wedge s} \over {u^3}}\hat \psi({\bf R}-{{\bf u} \over {2}}), \eqno
(2.13)$$
 where $\hat \psi $ if the Fourier transform of $\psi$
 $$ \hat \psi({\bf r})=\int {{d{\bf p}} \over {(2\pi)^{3 \over 2}}}\psi
 ({\bf p}){e^{i{\bf p.r}}}.\eqno (2.14)$$

One can thus answer questions such as: given the wave function of the photon
what is the mean value of the position observable? The answer includes
the usual expected term, the first term on the right hand side of
equation (2.13), and an additionnal term, absent for the non-relativistic
massive particle.

An important property of the Pryce
operator is that it has non-commuting components, the commutator
being given by
$$ [x_i,x_j]=-i\epsilon_{ijk}{p_{k}}{{\bf p.s}\over {p^3}}.\eqno (2.15)$$

\beginsection{3.Optimal localization}

The three components of the Pryce operator cannot be simultaneousely
diagonalized, so that one cannot find states localized exactly
at a given point in space. Newton and Wigner argued that this makes the
photon non-localizable. However, we will show that although the commutator
of the three components is non-vanishing, one can find states which are
"localised" within
an arbitrarily small region.

Before doing so we have to give a meaning to the statement "a photon
is localized in a
region (R) of space". For simplicity, take the region (R) to be
a ball surrounding the origin
$$ (R)\ \ \vec r ^{ \ 2}\leq r_0^2.\eqno(3.1)$$
We define a state to be localized in the region (R) if it is a
superposition of the eigenstates of the operator ${\bf x}^2$ with
eigenvalues less than or equal to $r_0^2$. Note that if  the components
of the
position operator were commuting, this definition would
agree with the usual one
that states that the state must be a superposition of the  eigenstates
of the position operator with eigenvalues in the region (R).

It turns out that the eigenvalues of the square of the Pryce operator
${\bf x}^2$ can be arbitrarily small but non-vanishing.
In momentum space, the solution to the equation
$$ {\bf x}^2\psi_{r_0}({\bf p})=r_0^2\psi_{r_0}({\bf p}) \eqno (3.2)$$
reads
$$ \psi_{r_0}({\bf p})=N\Psi(r_{0}{\bf p})\eqno (3.3)$$
where
N is a normalization constant and $\Psi$ is given by
$$\Psi({\bf u})={1 \over {\sqrt {u}}}J_{\sqrt{5}\over {2}}(u)E_{\pm }({\bf
u})v.
\eqno (3.4)$$
$J$ is the Bessel function, $v$ is an arbitrary constant vector and
$E_{\pm}$ is the projector onto positive and negative helicity states
$$ E_{\pm}={1 \over {2}}{{\bf u.s} \over {u}}  \left({ 1 \pm {{\bf u.s} \over
{u}}
 } \right).\eqno (3.5)$$

 We conclude
that one can find states which arelocalised, in the sense defined above, in the
region (R) with the radius $r_0$ having an arbitrary non zero value.

Multiplying the function $\Psi$ by $e^{-i{\bf p.x_{0}}}$ we translate the
origin to ${\bf x_0}$.

The function $\Psi$ vanishes at the origin. It is roughly speaking,
 a superposition
of momentum eigenstates corresponding to wavelengths smaller than $r_0$, from
this one can deduce an uncertainty relation, due to the non-commutativity
of the Pryce operator components
$$\Delta x \geq \lambda _{max} , \eqno (3.6)$$
$\lambda_{max}$ is the greater wavelength that contributes
to the photon wave function.

For photons  having a small dispersion in their wavelength, the usual
position momentum uncertainty dominates over the one given by the relation
(3.6).
 However, it would be interesting to see whether
 one can prepare experimentally photon states where
 the uncertainty due to non-commutativity
 of the position components dominates over the one due to the non-commutativity
 of the position and momentum operators. This would constitute a test for the
validity
 of the Pryce operator.

 \beginsection{4.Summary and conclusion}

  We showed that the only self-adjoint operator having
  the correct transformation properties
  under translations, rotations, parity and time inversion transformations,
  and that commutes with the projector over physical states, is the
  Pryce operator (2.11). We gave the expression of the mean
  value of the position observable in formula (2.13), and
   proved that one can find states which are localized in an arbitrarily
  small region in space. Finally, we argued that one of the experimental
  tests of the validity of this operator would be the test of the uncertainty
  relation (3.6) that arises due to the non-commutativity of
  the components of the Pryce operator
  .  Another test related to the Berry phase has been proposed by Skagerstam
   [10].

  \beginsection {Acknowledgements}

  I am grateful to H. Bergeron, T. Brodkorb, S. Caser and R. Omn\`es for many
  helpful discussions,
  and to
  A. Valance for encouragement.

  \beginsection {References}

  \item [1] C.K. Hong and L. Mandel, Phys. Rev. Lett. 56 (1986) 58.

  \item [2] T.D. Newton and E.P Wigner, Rev. Mod. Phys. 21 (1949) 400.

  \item [3] A.S. Wightman, Rev. Mod. Phys. 34 (1962) 845.

  \item [4] J.M. Jauch and C. Piron, Helv. Phys. Acta 40 (1967) 559.

  \item [5] W.O. Amrein, Helv. Phys. Acta 42 (1969) 149.

  \item [6] M.H.L. Pryce, Proc. Ror. Soc. (London) A195 (1948) 62.

  \item [7] H. Bacry, Annls. Inst. Henri Poincar\'e 49 (1988) 245.

  \item [8] E.P. Wigner, Ann. Math. 40 (1939) 149.

  \item [9] B.S. Skagerstam and A. Stern, Nucl. Phys. B294 (1987) 636.

  \item [10] B.S Skagerstam, preprint ITT 92-09,  (1992).

\end